\documentclass[smallextended]{svjour3}

\usepackage{natbib}
\usepackage{color}
\usepackage{latexsym} % for \Box
\usepackage{tikz-cd}

\newcommand\bm[1]{\mbox{\boldmath $#1$}}

\def\npk{\bm{k}}
\def\npl{\bm{l}}
\def\npm{\bm{m}}
\def\npmbar{\bm{\overline{m}}}

\newcommand{\Lemaitre}{Lema\^{\i}tre}

\newcommand{\smfrac}[2]{{\textstyle{#1\over#2}}}
\def\CKP{Cartan-Karlhede procedure}
\def\d{{\rm d}}
\newcommand{\eqref}[1]{(\ref{#1})}

\makeatletter
\newcommand{\ms}{\noalign{\vspace{3\p@ plus2\p@ minus1\p@}}}
\@addtoreset{equation}{section}
\makeatother

\begin{document}

\title{Spacetimes with continuous linear isotropies I: spatial rotations}

\author{M. A. H. MacCallum}
\institute{M. A. H. MacCallum \at School of Mathematical Sciences,
  Queen Mary University of London, Mile End Road, London, UK.
\email{M.A.H.MacCallum@qmul.ac.uk}}

\maketitle
\begin{abstract}
    The weakest known criterion for local rotational symmetry (LRS) in
    spacetimes of Petrov type D is due to \cite{GooWai86}. Here it is
    shown, using methods related to the \CKP, to be equivalent to
    local spatial rotation invariance of the Riemann tensor and its
    first derivatives. Conformally flat spacetimes are similarly
    studied and it is shown that for almost all cases the same
    criterion ensures LRS. Only
    for conformally flat accelerated perfect fluids are three
    curvature derivatives required to ensure LRS, showing that Ellis's
    original condition for that case is necessary as well as
    sufficient.
\end{abstract}

\section{Introduction}

\cite{Ell67} introduced the concept of local rotational symmetry of
spacetime (LRS). He set out three definitions for which he then proved
equivalence in the case of spatial rotations in dust spacetimes in general
relativity. His definition (B) reads ``At each point $P$ in an open
neighbourhood V of a point $P_o$, there exists a nondiscrete isotropy
group.'' Because only dust matter content was considered, this
isotropy had to be a spatial rotation group.
Assuming that the dust was not both shear-free and nonrotating, 
% statement before Lemma 4.1, p 1178.
Ellis proved the equivalence of (B) with the
definitions:\\
``(A${}_m$) At each point $P$ in an open neighbourhood
$U$ of a point $P_o$, there exists a nondiscrete subgroup $g$ of the
Lorentz group in the tangent space $T_P$ which leaves invariant the
curvature tensor and all its covariant derivatives to the $m$-th order''
(Ellis took $m=3$)\footnote{The notation has been altered here to avoid
  the same letter being used for different quantities.} and\\
``(C) There exists a local group of motions $G_r$ in an open
neighbourhood $W$ of a point $P_o$ which is multiply transitive on some
$q$[-dimensional] surface through each point $P$ of $W$''. (This
implies $r > q$).

\cite{SteEll68} then generalized the dust result to
spacetimes containing perfect fluid and/or an electromagnetic field.

This is the first of a set of papers finding the minimal $m$
in (A${}_m$) which guarantees (C) [and (B)] for the three cases where the
group $g$ contains respectively a spatial rotation, a boost and a null
rotation. For clarity $g$ will be referred to as a local invariance,
or a linear isotropy as defined below, `isotropy' without a preceding
qualifier being reserved for the spacetime isometry at each point $P
\in W$ which leaves $P$ fixed. When there is an isotropy, it induces a
local invariance: these papers address the converse, with the aim of
facilitating identification of such solutions (via the methods
discussed in Section \ref{sec:CKP}) if rediscovered in unusual
coordinates or from other assumptions.

Where there is isotropy, the spacetimes, following Ellis, will be
called locally rotationally symmetric (LRS). Analogously, `locally
boost symmetric' (LBS) and `locally null rotation symmetric' (LNRS) may
be used. LRSI, LBI and LNRI may similarly be used for the local invariances
imposed by (A${}_m$).

In the definition (A${}_m$), it is implicit that $g$ is the same at
all $P$ in $U$, i.e.\ that $g_P \cong g_{P_o}$ and $g_P$ at $P$ and
$g_{P_o}$ at $P_o$ have the same action on the curvature and its
derivatives at those points. The spacetime metric is also assumed to
be of class $C^{n},~n\geq m+2$, so that the $m$-th covariant
derivatives of the curvature are well defined and continuous.
% (A${}_m$) for a suitable $m$ will be assumed
% for each case studied: when
(C) $\Rightarrow$ (A${}_m$) for all $m \leq n$. For brevity,
(A${}_\infty$) will mean that (A${}_m$) is true for all $m$ for
which the derivatives are continuous, but that notation does not mean
that the metric is assumed to be $C^\infty$.

The elements of $g$ form the linear isotropy group at $P$: here
`linear' distinguishes a group of transformations of the tangent plane
at $P$ from a group of motions in the spacetime, or a neighbourhood thereof,
leaving $P$ fixed (isotropies). The linear isotropies could also be
called local isotropies but to avoid confusion that term will not be
used in this paper: `local' here will mean that the same condition
holds throughout a neighbourhood. 

`Spacetime' here simply means a four-dimensional Lorentzian
manifold. The field equations of general relativity will not be used,
but cases where the Ricci tensor takes the form that would be implied
by specific matter content in general relativity will be referred to.

In \cite{Sik76}, \S 1.9, it is stated that any spacetime can be
characterized by second and third order invariants of the metric (the
Riemann tensor and its first derivatives): while this result, and 
hence the proof claimed by \citeauthor{Sik76}, are now known to be
false, as shown by \cite{KarAma82} and by the series of examples
\citep{Kou92,Ske97,Mac98,Ske00} needing higher derivatives,
culminating in the proof \citep{MilPel09} that for a particular metric
one needs the 7-th derivative of the Riemann tensor (a 9-th order
invariant of the metric), it provided the initial conjecture that
(A${}_1$) would be sufficient to imply isotropy, which will be proved
for LRSI or LBI Petrov type D spacetimes and almost all of the conformally
flat cases. This will be referred to as Siklos' conjecture.

Only Petrov type D and conformally flat spacetimes have a Weyl tensor
that can satisfy (A${}_0$) (or (A${}_m$) for larger $m$) with a linear
isotropy group $g$ containing spatial rotations and/or
boosts. Similarly, only Petrov type N or conformally flat spacetimes
can satisfy (A${}_m$) with null rotations in $g$. The boost and null
rotation cases are studied in the later papers of this series. The
spatial rotation and boost cases are related by the asterisk operation
of the GHP formalism \citep{GerHelPen73}.\footnote{The calculations
  for type D spacetimes with boost invariance, and for conformally
  flat spacetimes with spatial rotation invariance, were guided by
  this correspondence, using prior calculations for type D spacetimes
  with rotational invariance and conformally flat spacetimes with
  boost invariance.}  Note that both rotational and boost isotropies
lead to calculations, in an appropriate null tetrad $(\npk,\,
\npl,\,\npm,\,\npmbar)$, with a natural interchange symmetry between
$\npk$ and $\npl$.

There are rather more Ricci tensor types than Weyl tensor types that
can satisfy (A${}_0$) for spatial rotations (or boosts or null
rotations). For (A${}_0$) to apply to the whole Riemann tensor, the
Weyl and Ricci tensors must of course be appropriately aligned, and
there may be a nonzero Ricci scalar $R$.  The tracefree part of
the Ricci tensor can be characterized by its Segre type. The possible
isotropy groups of the Ricci tensor were listed by Segre type in Table
5.2 of \cite{SteKraMac03}. Table 1 here lists those with nontrivial
invariance groups\footnote{The notation here follows that in
  \cite{SteKraMac03}.} $\hat{I}_0$ which include a spatial rotation.
The possible $\hat{I}_0$ appear in the list of subgroups of the Lorentz
group in Table 6.1 of \cite{Hal04}, which follows work of Shaw
and Schell.
\begin{table}[ht] \tabcolsep0.7ex
  \centering\caption{Possible invariance groups of the Ricci tensor
    containing spatial rotations, by Segre type.}
  \vspace{2ex}
\label{tab:5.2}
\begin{tabular}{@{}ll@{}} \ms \hline\hline \ms
Invariance group & Segre type of the Ricci tensor \\ \ms\hline \ms
\ms
Spatial rotations in a plane & $[(11)1,1],[(11), Z\overline{Z}],[(11),2]$ \\
% Boosts & $[11(1,1)]$ \\
Boosts and rotations & $[(11)(1,1)]$ \\
% One-parameter group of null rotations & $[1(1,2)],[(1,3)]$ \\
Null rotations and spatial rotations & $[(11,2)]$ \\
% $SO(2,1)$: three-dimensional Lorentz group & $[1(11,1)]$ \\
$SO(3)$: rotations in three spatial dimensions & $[(111),1]$ \\
Full Lorentz group & $[(111,1)]$ \\ \ms
\hline\hline
\end{tabular}
\end{table}

\cite{CahDef68} showed that for all LRSI or LBI Petrov type D
spacetimes (A${}_2$) was a sufficient criterion for the spacetime to
be LRS or LBS. Subsequently \cite{GooWai86} gave criteria for the LRS
Petrov type D case in terms of properties of the values of the spin
coefficients and curvature expressed in a Newman-Penrose (NP) null
tetrad.

The methods used in this paper are introduced in Section 2.  They are
based on the \CKP\ for characterizing spacetimes and testing their
equivalence, and the software CLASSI embodying it using the
Newman-Penrose formalism. (Where CLASSI is mentioned below, hand
calculation, or suitable alternative software, could of course be
used.) The \CKP\ provides a natural context for the investigation of
(A${}_m$). With these methods it is shown that Ellis's original
characterization for LRS can be sharpened in almost all cases, and the
characterization of Goode and Wainwright re-formulated.

In Section 3, the Goode and Wainwright criteria are shown to be
equivalent to (A${}_1$), and to imply (A${}_m),~m\leq 4$, with a spatial
rotation as the local invariance. From the \CKP\ as outlined in Section
2, this implies that the spacetime is LRS and there is a local
isometry group $G_r~(r\geq 3)$. Thus the Siklos conjecture is proved
for the type D LRS case.

Many of the steps in Section 3 are not original, but
simply echo those of Goode and Wainwright. They are set out here
because they prompted and illustrate the strategy used in the rest of
this paper and its companion papers.  Having imposed conditions on
curvature consistent with the invariance being studied, one uses the
Bianchi identities and the conditions implied by (A${}_1$) in order to obtain
conditions on the spin coefficients, and then uses the Ricci
identities (``NP'' equations), any additional conditions implied by
(A${}_m),~m > 1$, and in some cases the commutator relations, to
complete the arguments.

Section 4 considers the (A${}_m$) required for LRS in LRSI conformally
flat spacetimes, including the cases studied by \cite{Ell67} and
\cite{SteEll68} not included in Section 3. Here a three-dimensional
group of spatial rotations is possible, not just the one-dimensional group
allowed by Petrov type D, provided the Ricci tensor has
Segre type [(111),1] or [(111,1)].

A somewhat different approach to the LRS perfect fluid cases (Segre
type [(111),1]), using a 1+3 ``threading'' description of spacetime,
was given by \cite{VanEll96}. In other related work, \cite{Mar97} and
\cite{MarBra99} considered the construction of the metrics for LRS
perfect fluid solutions starting from data on the curvature, some of
the rotation coefficients and expressions involving invariantly
defined coordinates. A large number of papers have considered solving
the field equations of general relativity for the LRS cases with
various matter contents: see for example sections 13.4 and 14.3, and
Chapters 15 and 16, in \cite{SteKraMac03}, and sections 2.6, 2.11 and
7.5 and Chapter 4 of \cite{Kra97}.

A complementary problem of finding the possible metrics and their
curvatures assuming a multiply-transitive group acting on nonnull
hypersurfaces (i.e.\ assuming at least a $G_4$ rather than a $G_3$)
was addressed in \cite{Mac80}, using a method due to
\cite{Sch68,Sch71}.

Whenever there is a local isometry group $G_r$ in a
neighbourhood, the result of \cite{Hal89} implies that the
neighbourhood is isometric to a region of a spacetime with a global
isometry group $G_r$ or larger, provided that one has a
simply-connected connected smooth Hausdorff manifold with smooth
Lorentz metric. The smoothness is essential since, for example,
junction conditions allow one to match Schwarzschild and
Szekeres spacetimes at a spherical boundary \citep{Bon76}.

One can also consider analogous results for discrete rather than
continuous linear isotropy groups: a first result for such
isotropies was proved by \cite{Sch69}, and another by
\cite{DebMcLTar79}. In collaboration with Filipe Mena I have worked
further on Schmidt's and related cases
\citep{Men00,Men01,MenMac02}. It was in updating and completing that
work that the issues covered in this paper arose. There has also been
consideration of spacetimes where there are homotheties, rather than
isometries, with fixed points \citep{Hal88}.

\section{The \CKP}
\label{sec:CKP}
\subsection{The procedure and spacetime characterization}
The \CKP\ for characterizing spacetimes relies on Theorem 9.1 in
\cite{SteKraMac03}, which is Cartan's result as stated by Sternberg
and Ehlers. The current way to apply this result was introduced by
Karlhede \citep{KarAma79,AmaKar80,Kar80a,Kar80} and is set out in
Chapter 9 of \cite{SteKraMac03} and elsewhere. It consists of the
following iteration, where ${\cal R}^q$  denotes the set
$\{R_{abcd},\,R_{abcd;f} ,\,\ldots,R_{abcd;f_1f_2\cdots f_q}\}$ of the
components of the Riemann tensor and its derivatives up to the $q$th.

\begin{list}{}{\setlength{\itemsep}{0pt}\setlength{\parsep}{0pt}}
\item[1.]{Set the order of differentiation $q$ to 0.}
\item[2.]{If $q >0$, calculate the $q$th derivatives of the Riemann tensor.}
\item[3.]{Find the canonical form of ${\cal R}^q$.}
\item[4.]{Fix the frame as far as possible by this canonical form, and
  note the residual frame freedom. The group of allowed
  transformations is a linear isotropy group $\hat{I}_q$.\footnote{The
    notation $\hat{I}$ is adopted to distinguish between the linear
    isotropy group at $P$ and, if it exists, the isotropy group $I$,
    the subgroup of the full isometry group leaving $P$ fixed.} Let us
  denote $\dim \hat{I}_q$ by $s_q$. Necessarily $s_q \leq s_{(q-1)},
  \forall q>0$.}
\item[5.]{Find the number $t_q$ of independent functions of space-time
  position in ${\cal R}^q$ in the canonically chosen
  frame. Necessarily $t_q \geq t_{(q-1)}, \forall q>0$.}
\item[6.]{If the isotropy group and number of independent functions
    are the same as at the previous step, let $p+1 = q$ and stop; if
    they differ (or if $q$ = 0) increment $q$ by 1 and go to stage 2.}
\end{list}
The iteration of 2-5 above for a given $q$ will be referred to as step
$q$ of the \CKP.  The quantities computed, i.e.\ the components in
${\cal R}^p$ in a canonically chosen frame, are scalar invariants
defined by the curvature and its derivatives: they are called Cartan
invariants. One may also use functions of these invariants,
e.g.\ their ratios, which are also scalar invariants. In particular
these may be spin coefficients or Ricci rotation coefficients
evaluated in the canonically chosen frame, as mentioned below. Such
quantities can be called extended or derived Cartan invariants, and
for brevity here will just be called invariants.

The space-time is then characterized locally by the canonical form
chosen, the successive linear isotropy groups and independent function
counts, and the values of the non-zero invariants. It is
sometimes convenient to use the term ``$s$ sequence'' for
$(s_0,\,s_1,\,s_2,\ldots)$ and similarly for $t$.  The isotropy group
of the space-time will have dimension $s = s_p ={\rm
  dim}{\thinspace}\hat{I}_p$.  Since there are $t_p$ essential
space-time coordinates, clearly the remaining $4-t_p$ are ignorable so
the isometry group has dimension $\displaystyle{r = s + 4 - t_p}$ (see
e.g.\ \cite{Kar80a}).  Note that if $s \geq 1$, (A${}_{(p+1)}$) is
true, and vice versa.

Considering stage 4 of the above iteration, one may note that linear
isotropies imply that the basis vectors in the subspaces invariant
under their action can be determined only up to those
transformations. Spatial rotations about an axis and boosts act
respectively in spacelike or timelike two-planes; a position-dependent
boost or rotation (respectively) can be used in the orthogonal
two-planes to determine the frame as far as possible. This cannot
determine the frame uniquely, but at best up to the linear
isotropies. Similar remarks apply to the null rotation case.

In the calculations that follow, particular canonical choices of frame
are used, but the outcomes are not frame dependent, being statements
about geometric properties of the curvature and its derivatives. Those
properties would merely be harder to recognize in a randomly chosen
frame.

When a quantity $Q$ is invariant under the remaining allowed
transformations of a canonical frame determined by the curvature and
its derivatives, then its derivatives will share the relevant
(A${}_m$) isotropy and it can be included in the count of the
$t_i$. In what follows this will often be true of specific spin
coefficients.  Note that if the invariance applies only after a frame
choice at step $i$ then the derivatives of $Q$ have to obey the
isotropy only at steps $j\geq(i+1)$. In practice the restrictions on
derivatives of $Q$ often follow from (e.g.) the NP equations, without
invoking (A${}_{i+1}$), but this has not been systematically checked
for all instances.

A (pseudo-)Riemannian manifold is said to be curvature-homogeneous of
order $k$, denoted CH${}_k$, if $t_i=0,~\forall~ 0\leq i \leq k$, and
proper CH${}_k$ if $t_{k+1}>0$.  Spacetimes which are CH${}_k$ for all
$k$ are homogeneous. \cite{MilPel09} studied CH${}_k$ spacetimes when
finding the unique spacetime whose characterization requires the 7th
derivative of the Riemann tensor: it is a proper CH${}_2$ manifold. In
the process they found that there are no CH${}_3$ spacetimes and only a few
possible CH${}_1$ and CH${}_2$ cases. Potential CH${}_1$ and CH${}_2$
cases have to be eliminated at some points in the following arguments.

\subsection{Possible $(s_i,\,t_i)$ pairs in the \CKP}
\label{PossST}

There is an interplay between the $t_i$ and $s_{(i+1)}$ in the
\CKP. At step $(i+1)$, the $t_i$ independent functions from step $i$
will define $t_i$ linearly independent vectors, their
gradients. $\hat{I}_{(i+1)}$ must preserve those vectors.  If
$\hat{I}_{(i+1)}$ is a one-parameter local invariance group of spatial
rotations or boosts, it acts in a two-dimensional non-null subspace,
and the orthogonal subspace $W$ in which the $t_i$ gradients lie is
also two-dimensional and non-null. Thus $t_i\leq 2$ in this case.

If $\hat{I}_{(i+1)}$ is a one-parameter local invariance group of null
rotations preserving a null vector $\npk$, $t_i\leq 2$ is again
true. This follows because $\hat{I}_{(i+1)}$ also preserves a spatial
vector orthogonal to $\npk$, so there is only a two-dimensional space
in which the $t_i$ gradients can lie, although that space is not
orthogonal to the two-space in which the null rotation invariance
acts. Thus if (A${}_\infty$) holds with $s=1$ there is at least a
$G_3$ acting on the spacetime.

If ${\rm dim}{\thinspace} \hat{I}_{(i+1)}=2$, the linear invariance
group consists either of boosts and rotations in orthogonal
two-planes, so $t_i=0$, or of null rotations preserving a null vector
$\npk$ in which case $t_i\leq 1$. If ${\rm dim}{\thinspace}
\hat{I}_{(i+1)}=3$, $t_i\leq 1$ again follows. (The group
$\hat{I}_{(i+1)}$ will be one of SO(3), SO(2,1), or the group
generated by two null rotations preserving different null vectors and
a spatial rotation. See Table 5.2 of \cite{SteKraMac03} and Table
6.1 of \cite{Hal04}.)

\begin{figure}[ht]
  \caption{Possible pairs $(s,\,t)$ for spacetimes with local linear
    isotropies, and their evolution in successive steps of the
    \CKP. For the full description see the text of Section
    \ref{PossST}.}
\label{eq:diagram}
% \raisebox{-.5ex}{$1 \rightarrow \mathbb{Z}_2 \rightarrow$}
\begin{center}
\begin{tikzcd}
  (3,\,0)\arrow[r]\arrow[rd]\arrow[d] &
       (2,\,0) \arrow[r]\arrow[d]\arrow[rd]\arrow[ddr] & (1,\,0)\arrow[d] \\
  (3,\,1)\arrow[r] & (2,\,1) \arrow[r]\arrow[rd] & (1,\,1) \arrow[d]\\
            & & (1,\,2)
\end{tikzcd}
\end{center}
\end{figure}
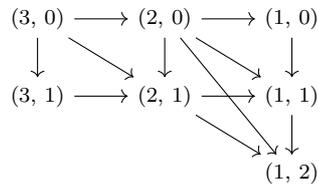
Figure \ref{eq:diagram} shows the resulting possible
successive pairs $(s,\,t)$ with $s_i\neq 0$ in the \CKP, assuming one
starts with $s_0\leq 3$ and ends with $s\geq 1$ (thus omitting the
possibility $s_0=6$ which leads only to the well-known conformally
flat Einstein spaces). The arrows join possible pairs at steps $i$ and
$i+1$ if the procedure does not terminate at step $i$.  The arrows
showing termination at step $i$, which are loops back to the same
values, have been omitted. Note that termination must happen at the
pair $(1,\,2)$ if $s=1$. Pairs $(s_i,\,t_i)$ and $(s_{(i+1)},\,
t_{(i+1)})$ may be joined by two or more arrows e.g.\ one could have
$(3,\,0)\rightarrow (1,\,2)$ in one step. There are of course
spacetimes for which $s=0$ but which have $s_i\geq 0$ for some $i<p$.
An example is given by the metric of
\cite{Kou92} that requires 4 derivatives ($p=3$) for its
characterization\footnote{I am grateful to Jan {\AA}man for drawing this
example to my attention.}: there
the $(s,\,t)$ sequence is  $(3,\,0)\rightarrow (1,\,1) \rightarrow
(0,\,3)\rightarrow (0,\,4)$.

\subsection{Implementation}
The \CKP\ is embodied in the software CLASSI which is used below.  A
more extended description of the \CKP\ was given by
\cite{PolSkedIn00a,PolSkedIn00b,PolSkedIn00c}, who provided modified,
corrected and extended details, as compared with CLASSI's version,
which were used in their Maple implementation. Note that the details
for the null rotation case are further corrected by the results of
\cite{Mac20}.

CLASSI was developed by Jan {\AA}man and collaborators, using and
extending Inge Frick's SHEEP software, and is described in
\cite{MacSke94} and more briefly in \cite{Mac18} (see also the manual
\citep{Ama87} supplied with the software).  To carry out the
Cartan-Karlhede procedure CLASSI uses the Newman-Penrose formalism as
set out in Chapter 7 of \cite{SteKraMac03}\footnote{Except that here
  the prime on the second index of $\Phi_{AB'}$ is retained and the NP
  scalar $\Lambda=R/24$ is used in preference to $R$. Note that in an
  Einstein space this is a multiple of the cosmological constant,
  usually also denoted by $\Lambda$. Calling this $L$ for clarity,
  $L=R/4$.}.  The ``Newman-Penrose equations'' (the Ricci equations)
and Bianchi identities [(7.21a)-(7.21r) and (7.32a)-(7.32k) in
  \cite{SteKraMac03}] will be referred to below as (NPa)--(NPr) and
(Ba)--(Bk).  For any scalar $Q$, the vector defined by $Q_{,a}$ is\\
\begin{equation}\label{NPderivs}
  Q_{,a}(2m^{(a\!\phantom{)}}\bar{m}^{\phantom{(}\!b)}
 -2k^{(a\!\phantom{)}}l^{\phantom{(}\!b)}) = \delta Q\,
 \npmbar+\bar{\delta}Q\, \npm - DQ\, \npl - \Delta Q\, \npk.
\end{equation}

As implemented in CLASSI, the \CKP\ uses totally symmetrized
multi-index spinors.  A shorthand notation for those will be used
here: it first appeared in print in \cite{KarAma80}.  It is an
extension of the Newman-Penrose notation, in which, for example, the
completely symmetric Weyl spinor $\Psi_{ABCD}=\Psi_{(ABCD)}$ is
represented by 5 components\footnote{This use of upper case Latin
indices for two distinct but related purposes does not lead to
confusion in practice.} $\Psi_{A},~A=0\ldots 4$, formed by
contractions with the basis spinors $(o^B,\,\iota^B)$.  The index $A$
on $\Psi_{A}$ counts the number of contractions with $\iota^B$: thus,
for example, $\Psi_{3}=\Psi_{(BCDE)}o^B\iota^C\iota^D\iota^E$.

This notation is extended as follows. Suppose that
$Q^{BCD\ldots}{}_{E'F'\ldots}$ is a completely symmetric multi-index
spinor, so that
$Q^{BCD\ldots}{}_{E'F'\ldots}=Q^{(BCD\ldots)}{}_{(E'F'\ldots)}$. (Here
the unprimed indices have been raised to enable clarity in the use of
the standard notation for symmetrization, but the shorthand notation
actually assumes all are covariant, i.e.\ lowered, in order to avoid
the sign issues arising because for any spinors $S^AT_A=-S_AT^A$.)
Such a spinor is said to have valence $(m,\,n)$ if it has $m$ unprimed
and $n$ primed indices. Then $Q_{AB'}$ denotes the component in which
$A$ of the unprimed indices and $B$ of the primed indices are
contracted with the basis spinors $\iota^A$ and $\bar{\iota}^{W'}$
respectively (and the other indices with $o^A$ and $\bar{o}^{X'}$). As
an example, consider the totally symmetrized part of the second
covariant derivative of the Weyl spinor,
$\nabla_{AA'}\nabla_{BB'}\Psi_{CDEF}$. It has valence $(6,\,2)$. The
component $\nabla^2\Psi_{41'}={\nabla^{\left( A \right.}}_{\left(
  X'\right.}{\nabla^B}_{\left. W'\right)}\Psi^{\left. CDEF\right)}
o_Ao_B\iota_C\iota_D\iota_E\iota_F\bar{o}^{X'}\bar{\iota}^{W'}$.

A minimal set of Cartan invariants in the form of totally symmetrized
spinors, taking into account their interrelations, and sufficient for
the above procedure, was defined by \cite{MacAma86}. The set defined
there contains the totally symmetrized derivatives of $\Psi$, $\Phi$
and $\Lambda$, together with, at order 1,
$\Xi_{DEFW'}={\nabla^C}_{W'}\Psi_{CDEF}$ and at order $q+2 \geq 2$,
the d'Alembertians of quantities at order $q$. Although in principle
unnecessary, it may be convenient at step $q$ to consider in addition the
gradients of quantities at step $(q-1)$.

For an LRSI totally symmetric spinor $Q$ of valence $(m,\,n)$ only the
components $Q_{AB'}$ with $2(A-B)=m-n$ are invariant under spatial
rotation in the $(\npm,\,\npmbar)$ plane; thus all other components
must be zero. If the spacetime is LRSI, a canonically chosen frame will
be one in which this is true. 

\section{Criteria for LRS Petrov type D spacetimes}
\label{sec:LRSD}

It is known that in general Petrov type D vacua need at most 3 steps
for classification \citep{ColdInVic90}: in the non-vacuum case an upper
bound of 6 steps was found \citep{ColdIn93}. Here it is shown that the
smaller bound $p=1$ applies when LRSI is assumed.

Petrov type D Weyl tensors are invariant under both boost and spatial
rotation isotropies. If the Ricci tensor shares those symmetries, $s_0=2$
and the only possible non-zero component of $\Phi_{AB'}$ in a
canonical null tetrad is
$\Phi_{11'}$. If $s_1=2$, so that $\hat{I}_1$ consists of both boosts
and rotations, it has no fixed vectors so invariance of first
derivatives of the curvature implies $t_0=0$. None of the first
derivative symmetrized spinor quantities defined by \cite{MacAma86} is
invariant under both boosts and rotations, so they must all be
zero; hence $t_1=0$ and the \CKP\ terminates at the first step. The
resulting homogeneous spacetimes, (12.8) in \cite{SteKraMac03},
consist of two orthogonal two-dimensional spaces of constant curvature
and admit an isometry group $G_6$, like the well-known
Bertotti-Robinson solution\footnote{This form was actually first
  studied by Levi-Civita in 1917: see \cite{SteKraMac03}, Section
  12.3.}.

The remaining Petrov type D cases with $s>0$ must have $s_1=1=s$, so
$\hat{I}_1$ consists of either rotations or boosts but not both.  Here
the case with rotations is studied, and in a companion paper the case
with boosts.  By the argument in Section 2, $t_i \leq 2$ at all $i\geq
0$ in either case. Thus the \CKP\ must terminate at $p\leq 3$, the
worst case sequence being
$(2,\,0)\rightarrow(1,\,0)\rightarrow(1,\,1)\rightarrow
(1,\,2)\rightarrow (1,\,2)$: in this case the spacetime is
proper CH${}_1$. Thus showing that (A${}_4$) holds will always suffice to imply
LRS for Petrov D spacetimes with LRSI.

Since the Petrov type is assumed to be D, a Newman-Penrose null tetrad
can be chosen which is invariant under the group of spatial rotations
in the $(\npm,\,\npmbar)$ plane and in which the only nonzero $\Psi_A$ is
$\Psi_2$. For LRSI the Ricci tensor and the first derivatives
of curvature must also be invariant under those rotations.
% File lrsd.nul and lrsd.log
The rotational invariance of $\nabla \Lambda$ implies 
\begin{equation}
\label{dlambdlrs}
\delta \Lambda = 0.
\end{equation}
Of the $\Phi_{AB'}$ only $\Phi_{00'}$, $\Phi_{11'}$ and $\Phi_{22'}$
can be nonzero.   Rotational
invariance implies that the gradient of the scalar $\Psi_2$ in the
symmetry plane is zero so $\delta\Psi_2=0
=\bar{\delta}\Psi_2$. Rotational invariance of the Cartan invariants
$\nabla{\Psi_{AB'}}$ requires that
 $$\nabla{\Psi_{01'}} = \nabla{\Psi_{11'}} = \nabla{\Psi_{21'}} =
 \nabla{\Psi_{30'}} = \nabla{\Psi_{40'}} =\nabla{\Psi_{41'}} =0, $$
which give respectively
% File gtyped.nul
\begin{equation} \label{C1values} \kappa=\sigma=\tau=\pi=\lambda=\nu=0.
\end{equation}
The possibly non-zero components are $\nabla\Psi_{20'}$ and
$\nabla\Psi_{31'}$. (The components not mentioned above are identically zero.) 
One may note that the results for $\kappa$, $\sigma$, $\lambda$ and
$\nu$ show that condition (c) of the  Kundt-Thompson theorem (Theorem
7.5 in \cite{SteKraMac03}) is satisfied, and that \eqref{C1values} implies
rotational invariance of $\Psi$.

Thus in Petrov type D invariance of the Riemann tensor and its first
derivatives under spatial rotation implies that there is a
Newman-Penrose tetrad (a canonical one for Petrov type D, unique up to
a spatial rotation and boost) in which the following criteria hold
\begin{eqnarray}
  {\rm (C1)}: &&  \kappa=\sigma=\tau=\pi=\lambda=\nu=0.\\
  {\rm (C2)}: &&  \Phi_{01}=\Phi_{12}=0.\\
  {\rm (C3)}: &&  \delta \Lambda=0.
\end{eqnarray}
\cite{GooWai86} proved:
\begin{theorem}\label{thm2.1}{\rm(Theorem 2.1 of \cite{GooWai86})}
A space-time $($assumed conformally curved$)$ is LRS if
and only if there exists a null tetrad $(\npk,\,
  \npl,\,\npm,\,\npmbar)$ in which
{\rm (C1)-(C3)} hold.
\end{theorem}

Conversely, (C1)--(C3) imply local rotational invariance of the
curvature and its first derivatives.  For the curvature this requires
that $\Phi_{02'}=0$, which follows immediately from (C1) and (NPg).
For the first derivatives $\nabla \Lambda$ and $\nabla\Psi$ it has
already been shown that LRSI follows from (C1) and (C3).
The components of $\nabla\Phi$ that must vanish for LRSI are given in full
as \eqref{DPhi01}--\eqref{DPhi23} in Section \ref{CFcases}, and the
relations between those components of $\nabla\Phi$ and the Bianchi
identities are discussed there. Given (C1), those components all vanish, due to
(Ba) and (Bd).

From (Bj) with (C1)--(C3) $\delta\Phi_{11'}=0$. The LRSI of the
gradients of the other admissible components of the Ricci curvature
($\Phi_{00'}$ and $\Phi_{22'}$) is equivalent to
$(\bar{\alpha}+\beta)=0$, from (Ba) and (Bd). To prove from (C1)--(C3)
that this holds requires some calculation.  The detailed arguments of
\cite{GooWai86} will be repeated here because they illustrate the
strategy to be used in other cases and because variants of them also
work for the conformally flat cases to be considered later (and for
the local boost invariance to be considered in a companion paper).

\cite{GooWai86} begin their proof by substituting (C1) in (NPb, c,
g, i and p) and using (C2) to show that
$$\Phi_{02}=\Psi_0=\Psi_1=\Psi_3=\Psi_4=0.$$ Therefore the spacetime
must have Petrov type D or 0 and its Riemann tensor is invariant under
the rotations.  (Bg) and (Bh) now give
$\delta\Psi_2=\bar{\delta}\Psi_2=0$.

At this point the frame is only fixed up to a position-dependent boost
(and an arbitrary spatial rotation). Under a boost with parameter $A$
and a spatial rotation through $\theta$, (C1)--(C3) are preserved and
$(\bar{\alpha}+\beta)^\star= [(\bar{\alpha}+\beta)+\delta
  A]e^{i\theta}$, where $\star$ denotes the transformed value.
\citeauthor{GooWai86} use the NP equations, Bianchi identities and the
commutator expressions to show that one can choose a boost to obtain a
frame in which\footnote{It is this possibility that \cite{CahDef68}
  overlooked, instead using invariance of second derivatives to obtain
  their equivalent result: for some details see the Appendix. Note
  that use of the commutators implies the existence but not the
  invariance of second derivatives.}
\begin{equation}\label{aplusb}
  \bar{\alpha}+\beta =0.
\end{equation}
This will complete the equivalence of (C1)-(C3) with an assumption of
rotational invariance of the Riemann tensor and its first derivative
in a Petrov type D spacetime. (The rotational invariance of
$\Xi_{AB'}$ is readily checked.) So all first derivatives of curvature
including $\nabla\Phi$  will then be LRSI.

To prove that (C1)--(C3) imply \eqref{aplusb} and the further
equations which will eventually imply that (A${}_4$) is manifestly true,
one first notes that (Bk) minus (Bf) gives
\begin{equation}\label{DPhiR}
\Delta(\Phi_{11'}+\Lambda-\Psi_2) =
\rho\Phi_{22'}+3\mu\Psi_2-2\bar{\mu}\Phi_{11'}.
\end{equation}
The commutator
$[\delta,\,\Delta]$ is applied to this. (Bd, h and j) yield
$\delta(\Phi_{11'}+\Lambda-\Psi_2)=0$, and (NPk and m) give
$$ \delta\rho=(\bar{\alpha}+\beta)\rho \quad {\rm and} \quad
\delta\bar{\mu}=-(\bar{\alpha}+\beta)\bar{\mu}.$$ Using these, the commutator
gives
$$\delta\mu=-(\bar{\alpha}+\beta)\mu$$ if $\Psi_2\neq 0$.  Thus if
$\mu\neq 0$, boosting the frame by $A=\sqrt{\mu\bar{\mu}}$ will give
\eqref{aplusb}. Similarly applying $[\bar{\delta},\,D]$ to the
difference of (Bg) and (Bi), which gives
\begin{equation}\label{DelPhiR}
  D(\Phi_{11'}+\Lambda-\Psi_2)=2\bar{\rho}\Phi_{11'}-3\rho\Psi_2
  -\mu\Phi_{00'},
\end{equation}
one obtains
$\delta\bar{\rho}=\bar{\rho}(\bar{\alpha}+\beta)$ if $\Psi_2\neq
0$. This calculation analogously uses (Ba, g and j) and (NPk and m)
and shows that if $\mu=0$ one can boost by $A=\sqrt{\rho\bar{\rho}}$ to
obtain \eqref{aplusb}.

If both $\mu$ and $\rho$ are 0, calculating $[\delta,\,\bar{\delta}]A$
shows that the equations $\delta A = -A(\bar{\alpha}+\beta)$ and its
complex conjugate are compatible and thus have a solution, again
giving \eqref{aplusb}. 

To show the spacetime must be LRS, first substitute from
\eqref{aplusb} into (NPd) and (NPe) to show $\delta(\varepsilon
+\bar{\varepsilon})=0$ and similarly, from (NPo) and (NPr),
$\delta(\gamma+\bar{\gamma})= 0.$ (One may note that $\rho$, $\mu$,
$(\varepsilon+\bar{\varepsilon})$ and $(\gamma+\bar{\gamma})$ are
invariant under the spatial rotations.)

Summarizing, \eqref{C1values} implies that in a frame with boost
chosen to ensure \eqref{aplusb} the following hold.
\begin{equation}\label{deltas}
  \delta\rho=\delta\bar{\rho}= \delta\mu=\delta\bar{\mu} =
  \delta{\Phi_{00}}=\delta\Phi_{22}=\delta(\varepsilon
  +\bar{\varepsilon})=\delta(\gamma+\bar{\gamma})= 0.
\end{equation}

In both \cite{CahDef68} and \cite{GooWai86} the proof that the
spacetime is LRS is completed by introducing coordinates and finding
the possible metrics. However, one can obtain that result in a
coordinate-free manner using Cartan invariants. Inserting the
equations \eqref{C1values}, \eqref{aplusb} and \eqref{deltas} into the
spinor $k$-th derivative quantities of \cite{MacAma86} for $k=1\ldots
4$, one finds (A${}_4$) holds.  (The further equations between spin
coefficients etc found by \cite{GooWai86} are not needed in obtaining
this result, though they may simplify the final expressions obtained.)
Thus the \CKP\ must terminate\ at $p=3$ or earlier, with  $s\geq 1$
and $t_p\leq 2$. This gives:

\begin{theorem}\label{LRSthm}
  If a neighbourhood of a spacetime of Petrov type D is such that the
  Riemann tensor and its first derivative are invariant under a local
  spatial rotation isotropy, then the spacetime is locally
  rotationally symmetric and admits a multiply-transitive $G_r ~(r\geq
  3)$.
\end{theorem}

The converse is of course trivial.

One can show that in fact (A${}_3$) is sufficient to prove Theorem
\ref{LRSthm}, i.e.\ that the spacetime cannot be proper
CH${}_1$, as follows.\footnote{\cite{MilPel09} have shown there are no such
  spacetimes for type D. This is also implied by the simple proof
  which follows and shows that for a $t$ sequence with $t_2=1$,
  $t_1=t_0=0$ the \CKP\ terminates with $p\leq 2$.}

% File lrsd0.nul
If $t_0=0$, $\Psi_2$ must be constant and constancy of
$\nabla\Psi_{AB}$ then implies that $\rho$ and $\mu$ are constant. Any
non-zero $\Phi_{AB'}$ are also constant. If at least one of $\rho$ or
$\mu$ is non-zero then, from (NPa and q), or (NPh and n),
$\varepsilon+\bar{\varepsilon}$ and $\gamma+\bar{\gamma}$ are constant
(possibly zero).  Calculation (using CLASSI) then shows that $t_2=0$
i.e.\ all terms in the second derivatives would also be constant, so
the \CKP\ would terminate at step 1 or 2.  If both $\rho$ and $\mu$
are zero, (NPa) and (NPn) imply $\Phi_{00'}=\Phi_{22'}=0$, and then
inspection (using CLASSI) shows that all first derivatives of the
Riemann tensor, and hence all higher derivatives, are zero, and the
\CKP\ would terminate at step 1.

In showing that (A${}_4$) [{\em a fortiori}, (A${}_3$)] is true, in
order to prove Theorem \ref{LRSthm}, the rotational invariance of the
Ricci tensor and its derivatives up to the fourth was checked but no
assumption on invariance of these quantities other than (C2)--(C3) was
made. In the conformally flat case one usually has to assume, rather
than being able to derive, $\Phi_{02'}=0$. The nonvanishing of
$\Psi_2$ in Petrov type D was used in deriving \eqref{C1values} and
\eqref{aplusb} but was not otherwise required in proving
(A${}_4$). \eqref{C1values} and \eqref{aplusb} imply \eqref{deltas}
due to the NP equations. Thus if one can obtain \eqref{C1values} and
\eqref{aplusb} in a conformally flat spacetime whose Ricci tensor is
rotationally invariant, (A${}_4$) will hold and the spacetime will be
LRS.  The arguments in the Petrov D case suggest that to achieve this
one must exploit the information in Ricci and Bianchi identities as
much as possible.

\section{Criteria for conformally flat spacetimes}
\label{CFcases}

The conformally flat cases to be considered have a Ricci tensor of one
of the Segre types appearing in Table 1.  (A${}_m$) is assumed to hold
with a group $g$ which contains  spatial rotations about at least one axis.

For Ricci tensors of Segre type [(111,1)], the linear isotropy
$\hat{I}_0$ is the whole Lorentz group. Only $\Lambda$ can be a
non-zero curvature component. The Bianchi identities show that
$\Lambda$ is constant, giving (locally) the well-known spacetimes of
constant curvature: Minkowski spacetime and the de Sitter and anti de
Sitter spacetimes. The subgroup $g$ in (A${}_m$) is the trivial one
comprising the whole Lorentz group. So $s=6$, $t_p=0=t_0$, and there
is a group $G_{10}$ transitive on the whole spacetime. In these cases
the \CKP\ terminates at the first step and (A${}_0$) suffices because
it will imply (A${}_1$).

The rest of this section studies the (A${}_m$) required for LRS in
conformally flat spacetimes with Ricci tensors of the Segre types in
Table 1 which are less special than [(111,1)]. Obviously $m\geq 1$ is
needed.

\subsection{The first derivatives of the Ricci tensor}
\label{CF1stDerivs}
% File cflri.nul
The first step is to impose LRSI on the first derivatives of $\Phi$
and $\Lambda$. Then one can try first to derive \eqref{C1values},
which was obtained in Petrov type D from invariance of $\nabla\Psi$,
and then \eqref{aplusb} and \eqref{deltas}.

In the Segre types that admit a spatial rotation (which is again
assumed to be in the $(\npm,\,\npmbar)$ plane) only the components of
$\Phi_{AB'}$ for which $A=B$ can be nonzero. So (C2) and
$\Phi_{02'}=0$ are assumed. For (A${}_1$) to hold with LRSI also
implies (C3). When the frame has been fixed sufficiently that components of
$\Phi$ become invariants, the $\delta$ derivatives of those components
must vanish.

The components of $\nabla\Phi$ that must vanish for LRSI are
\begin{eqnarray}
3\nabla\Phi_{01'}&=&
 \delta\Phi_{00'}-2(\bar{\pi}+\bar{\alpha}+\beta)\Phi_{00'}
 +4\kappa\Phi_{11'}, \label{DPhi01}\\
3\nabla\Phi_{02'}&=& 2(2\sigma\Phi_{11'}-\bar{\lambda}\Phi_{00'}),\label{DPhi02}\\
9\nabla\Phi_{12'}/2&=& (\kappa\Phi_{22'}-2\bar{\pi}\Phi_{11'})
                   +(2\tau\Phi_{11'}-\bar{\nu}\Phi_{00'}),\label{DPhi12}\\
3\nabla\Phi_{13'}&=&2(\sigma\Phi_{22}-2\bar{\lambda}\Phi_{11'})
 ,\label{DPhi13}\\
3\nabla\Phi_{23'}&=&
 \delta\Phi_{22'}+2(\tau+\bar{\alpha}+\beta)\Phi_{22'}-4\bar{\nu}\Phi_{11'}.\label{DPhi23}
\end{eqnarray}
Note that $\nabla\Phi_{03'}\equiv 0$.  Inspecting the Bianchi
identities, one finds that (Bb) implies \eqref{DPhi02}, (Bh) and the
conjugate of (Bg) imply \eqref{DPhi12}, and the conjugate of (Bc)
implies \eqref{DPhi13}. Subtracting (Ba) from \eqref{DPhi01}, and
subtracting the conjugate of (Bd) from \eqref{DPhi23} leaves
\begin{equation}\label{DPhiRem}
  \kappa\Phi_{11'}-\bar{\pi}\Phi_{00'}=0 \quad {\rm and} \quad
  \tau\Phi_{22'}-2\bar{\nu}\Phi_{11'}=0,
\end{equation}
as the only possible information from rotational invariance of
$\nabla\Phi$ not already contained in the Bianchi identities. Hence if
\eqref{DPhiRem} are satisfied $\nabla\Phi$ will be rotationally
invariant.  [In Segre type [(111),1] (``perfect fluid'')
  \eqref{DPhiRem} gives no new information, being equivalent to (Bg)
  and (Bh) in the canonical frame in which
  $\Phi_{00'}=\Phi_{22'}=2\Phi_{11'}$]. If (C1) holds, \eqref{DPhiRem}
is identically satisfied, but in some conformally flat cases
\eqref{DPhiRem} is needed in order to infer (C1).

\subsection{Conformally flat spacetimes with Ricci tensors of Segre
  types for which $s_0=1$}
\label{ssec:CFG1}

These Segre types appear in the first line of Table 1: they are types
[(11)1,1], [(11)$,Z\bar{Z}$], and [(11),2]. Taking the spatial
rotation to act in the $(\npm,\,\npmbar)$ plane, these have canonical
forms with $\Phi_{22'}\neq 0\neq \Phi_{11'}$ (and
$\Phi_{01'}=\Phi_{02'}=\Phi_{12'}=0$). The three different cases,
[(11)1,1], [(11),$Z\bar{Z}$], and [(11),2], are characterized by
$\Phi_{00'}\Phi_{22'}>0$, $\Phi_{00'}\Phi_{22'}<0$ and $\Phi_{00'}=0$
respectively. None have a generally accepted physical interpretation:
however, the combined perfect fluid and electromagnetic field
solutions of \cite{SteEll68} have type [(11)1,1].  Avoiding the more
special type [(111),1] requires that $\Phi_{00'}\Phi_{22'}\neq
4\Phi_{11'}^2$, which is assumed for the rest of this section. Since
$\Phi_{11'}$ is invariant under the remaining frame freedom (a
position-dependent boost),  (A${}_1$)  implies $\delta\Phi_{11'}=0$.

If these spacetimes obey (A${}_\infty$) the \CKP\ must terminate after
at most 3 steps ($p=2$) since $s=1$ at every step and  $t_2$
is at most 2 (since $t_0 \geq 0$) which is the maximum possible $t_p$. So
(A${}_3$) would suffice.  All the resulting spacetimes would again
admit a $G_r,~r\geq 3$ acting on submanifolds of dimension at least 2.

Assuming (A${}_1$) and using the Bianchi identities, one can now
obtain \eqref{C1values}. (Bb) and (Bc), or
equivalently \eqref{DPhi02} and \eqref{DPhi13}, imply that
$\sigma=\lambda=0$ (provided $\Phi_{00'}\Phi_{22'}\neq
4\Phi_{11'}^2$). Similarly \eqref{DPhiRem}, (Bg) and (Bh), which
read
$$ Bg:~ \bar{\kappa}\Phi_{22'}-2\pi\Phi_{11'}=0\quad {\rm and} \quad
Bh:~ 2\tau\Phi_{11'} -\bar{\nu}\Phi_{00'}=0.$$ give $\kappa=\pi=0$ and
$\tau=\nu=0$, completing the derivation of \eqref{C1values}.

If the remaining boost freedom is used in the cases [(11)1,1] and
[(11),$Z\bar{Z}$] to set $\Phi_{00'}=\pm\Phi_{22'}\neq 0$ then (Ba) and
(Bd) imply \eqref{aplusb}.  As in Section 3 this implies
$\delta(\varepsilon +\bar{\varepsilon})=0=\delta(\gamma+\bar{\gamma})=
0.$ From (NPi) and (NPm) one has $\delta\rho=\delta\bar{\mu}=0$. To
obtain the remaining equations in \eqref{deltas}, one can either argue
that since $\rho$ and $\mu$ are invariants under the allowed changes
of frame, $\bar{\delta}\rho=\delta\rho=\delta{\mu}=\bar{\delta}\mu=0$
or consider applying $[\delta, D]$ to $\Phi_{11'}+\Lambda$ using
\eqref{DPhiR} and $[\bar{\delta}, \Delta]$ to it using
\eqref{DelPhiR} to obtain a pair of linear equations for
$\bar{\delta}\rho$ and $\delta\mu$ leading to the same
conclusion. Then as in Section \ref{sec:LRSD} this ensures (A${}_4$)
and the spacetimes are LRS.

For Segre type [(11),2] one can use the allowed boost to make
$\Phi_{22'}=\Phi_{11'}\neq 0$ so that $\delta\Phi_{22'}=0$ and hence,
from (Bd), \eqref{aplusb} again holds. Then the rest of \eqref{deltas}
and the conclusion that the spacetime is LRS follow as before.

Thus for these cases of rotational invariance, Theorem \ref{LRSthm}
holds with `spacetime of Petrov type D' replaced by `conformally flat
spacetime'.

\subsection{Conformally flat spacetimes with Ricci tensors of Segre
  type   [(11)(1,1)]}

% File cf1111.nul (not the same one as in the null rotation case!)
Here $s_0=2$. Only $\Phi_{11'}$ and $\Lambda$ can be non-zero Ricci tensor
components, and both are invariant under rotations and thus obey
$\delta Q = \bar{\delta} Q=0$.  Assuming $\Phi_{11'}\neq 0$, (Ba) to (Bd) give 
$$\kappa=\sigma=\lambda=\nu=0.$$ If $s=2$ one must have $t_1=0$ (as the
boost and rotations together have no invariant vectors) and so the
\CKP\ terminates at step 1 since $s_1=s_0$ and $t_1=t_0$. Hence as in
Section \ref{sec:LRSD} one has the Bertotti-Robinson type spacetimes
with a $G_6$ transitive on the whole spacetime, and (A${}_1$)
suffices.

If $s=1$ and $\hat{I}_1$ is a group of spatial rotations then from
(Bg) and (Bh) $\pi=\tau=0$ so \eqref{C1values} holds.  To obtain
\eqref{aplusb}, note that (NPk) gives $\delta\rho =
\rho(\bar{\alpha}+\beta)$ and (Be) and (Bi) give
$D\Phi_{11'}=(-\rho+2\bar{\rho})\Phi_{11'}$. Applying the commutator
$[\delta,\,D]$ to $\Phi_{11'}$, gives $\delta\bar{\rho} =
\bar{\rho}(\bar{\alpha}+\beta)$. Similarly the conjugate of (NPm)
gives $\delta \bar{\mu}=-\bar{\mu}(\bar{\alpha}+\beta)$, (Bf) and (Bk)
give $\Delta\Phi_{11'}=(\mu-2\bar{\mu})\Phi_{11'}$, and applying
$[\delta,\,\Delta]$ to $\Phi_{11'}$ gives $\delta\mu=
-\mu(\bar{\alpha}+\beta)$. Then as in Section \ref{sec:LRSD} there is
a boost such that \eqref{aplusb} holds, and hence the spacetime is LRS.

Thus for a conformally flat spacetime with a Ricci tensor of Segre
type $[(11)(1,1)]$, (A${}_1$) with a one-dimensional group of spatial
rotations implies LRS.

\subsection{Conformally flat spacetimes with Ricci tensors of Segre type [(11,2)]}
\label{CFpure}

Here $s_0=3$, $\hat{I}_0$ being generated by a two-parameter group of
null rotations about $\npk$, say, and spatial rotations in a plane
orthogonal to $\npk$.  Only $\Phi_{22'}$ and $\Lambda$ can be non-zero
in a frame adapted to the null rotation isotropies. Such a Ricci
tensor can represent a null electromagnetic field or other forms of
``pure radiation''.  The conformally flat pure radiation solutions
with $\Lambda=0$ are completely known (Theorem 37.19
in \cite{SteKraMac03}). One set are plane waves, with an obvious
rotational symmetry, as well as null rotation symmetry ((37.104) in
\cite{SteKraMac03}), and the other set ((37.106) in
\cite{SteKraMac03}), the Edgar-Ludwig metrics, cannot have local rotational
symmetry (\cite{Ske97}).

 Note first that an $\hat{I}_i, ~i\geq 1$ with $s_i=2$ cannot
include both a spatial rotation (by an angle $\theta$ say) and a
one-parameter group of null rotations, with parameter $B=|B|e^{i\psi}$
say, because the spatial rotation would transform those null rotations
to a different one-parameter null rotation invariance group with
parameter $B'=Be^{i\theta}=|B|e^{i(\theta+\psi)}$. Thus if
$s_i=2,~i\geq 1$ the remaining isotropies are the two null rotations,
a case treated in a companion paper, and if $\hat{I}_p$ contains the
spatial rotation but $s_p\neq 3$, $s_1=1$. By the same logic as in
Section \ref{ssec:CFG1}, (A${}_3$) will suffice to imply LRS, the
\CKP\ then terminates at step 3 or earlier and there is a $G_r$,
$r\geq 3$ (which will act in a spacelike surface).

To obtain the minimal $m$ by the methods of this paper, a boost is
used to set $\Phi_{22'}=1$. The remaining frame freedoms are
position-dependent rotations.  One has $\delta \Lambda=0$ and the
Bianchi identities (Bc), (Bf) and (Bk) give
$\kappa=\sigma=D\Lambda=0$, $\bar{\tau}=2(\alpha+\bar{\beta})$,
$\Delta \Lambda=\rho$, and
$\rho=-2(\varepsilon+\bar{\varepsilon})$. Imposing (A${}_1$) implies,
from $\nabla\Phi_{23'}=0$, that $\tau+\bar{\alpha}+\beta=0$ so
$\tau=\alpha+\bar{\beta}=0$. Then (NPc), (NPk) and (NPp) give
$\rho\bar{\pi}=\delta\rho~[=\delta(\varepsilon+\bar{\varepsilon})]
=\rho\bar{\lambda}=0$.  Subtracting (NPo) from the conjugate of (NPr)
gives $\delta(\gamma+\bar{\gamma})=-\rho\bar{\nu}/2.$ 
The two cases $\rho=0$ and $\rho\neq 0$ will now be considered.

%File cfpur0.nul
If $\rho=0=\varepsilon+\bar{\varepsilon}$, then $\Lambda$ is
constant. (NPq) gives $\Lambda=0$ and (NPf) then yields
$D(\gamma+\bar{\gamma})=0$. Then by direct calculation (A${}_3$) holds
and the spacetime is LRS. So (A${}_1$) suffices to prove LRS in the
case. The metric this leads to is (37.104) in \cite{SteKraMac03}, in
which the Ricci tensor is interpretable as that of an electromagnetic
field.

% File cfpure.nul (not the one used with null rotations)
If $\rho\neq 0$, $\Lambda$ cannot be constant (in particular not
zero). (NPq) implies $\mu$ is real; it is invariant under spatial
rotations so $\delta\mu=0$.  $(\gamma+\bar{\gamma})$ is also a spatial
rotation invariant so $\delta(\gamma+\bar{\gamma})=0=\nu$.  Thus all
the equations \eqref{C1values}, \eqref{aplusb} and \eqref{deltas} are
satisfied, (A${}_1$) implies (A${}_3$), and the spacetime is LRS. A
manifestly LRS example of such a spacetime is given by
% File cfprex.nul % r^2 -> new r
$$\d s^2 = \smfrac{1}{3}r^2 u\d u^2 + \d u \d r + 2 r^2(\d x^2+\d
y^2);$$ here $\Psi_2=1$ and $\Lambda=-2u/3$.
%MM Watch out for signs. cfprex.lor gives R = 16u but antdes.dia gives
% 12 so sign seems to be opposite to usual, and I have assumed that a
% correct relation in my signature is LAMBD = - R/24
Such metrics are unlikely
to be of physical interest.

\subsection{Ricci tensors of Segre type [(111),1] (perfect
  fluid)}

In general relativity, a Ricci tensor of Segre type [(111),1] is
usually referred to as a perfect fluid whether or not it obeys any
standard equation of state.  Conformally flat perfect fluids have been
rather fully investigated (see Section 37.4.2 of \cite{SteKraMac03}
and references therein). They are all known (see Theorem 37.17 of
\cite{SteKraMac03} and the paragraph following it). They are more
naturally treated in a 3+1 or orthonormal tetrad formalism than the
null tetrad used so far here. The correspondence between the NP spin
coefficients and the Ricci rotation coefficients in the orthonormal
tetrad (ONT) formalism (as set out in \cite{Mac73}), some of which
coincide with components of the kinematic quantities as defined by
\cite{Ehl61}, is given in the appendix to \cite{GooWai86}. In this
section the ONT variables will appear where they enable briefer
exposition.

\cite{Bra86} showed that the third covariant derivative was needed to
complete the classification of one subset of the solutions, those with
a non-expanding fluid, implying that $p=2$ in the \CKP. He also
constructed the solutions by a method that in effect inverts the
\CKP. \cite{BarRow90} studied the possible isometry groups admitted by
these cases, and their results were later re-obtained using the
\CKP\ (computed in an orthonormal rather than null tetrad) by
\cite{Sei92}. \cite{Bar98} studied the isometry groups of the
expanding metrics (Stephani metrics). This section studies the $m$
required in (A${}_m$) for the subset of the solutions which admit LRSI
to be LRS.

Note that this will complete the re-investigation of the original
arguments that (A${}_3$) implies LRS \citep{Ell67,SteEll68}: the
Petrov type D cases are covered in Section \ref{sec:LRSD} and the
conformally flat perfect fluid plus electromagnetic field cases in
Section \ref{ssec:CFG1}. Both those cases needed only (A${}_1$).

To avoid confusion with the spin coefficients $\mu$ and $\rho$, $\xi$
is used here for the fluid energy density (and $p$ for
pressure despite its different use in the \CKP). Then $R=\xi-3p$ and
in a frame such that the fluid flow vector ${\bm u} =
(\npk+\npl)/\sqrt{2}$ (fixing the frame up to swaps of axes and
spatial rotations), $\Phi_{00'}=\Phi_{22'}=2\Phi_{11'}=(\xi+p)/4$.
This is nonzero as otherwise the Segre type is $[(111,1)]$,
considered above.

The invariance group of the Riemann tensor is SO(3). Because that has no
two-dimensional subgroup, either $s_p=3$ or $s_p=1$. Consider first
the case $s_1=3$, i.e.\ assume (A${}_1$) with SO(3) as $g$. This
leads only to the well-known Robertson-Walker geometries\footnote{In general
  relativity, called Friedman-\Lemaitre-Robertson-Walker spacetimes in
  recognition of the pioneering work on their dynamics.}, a result
% Also in Ehl61 referring to Robertson?
stated in this form by \cite{Ell67} for dust (and repeated for a more
general perfect fluid in \cite{SteEll68}).

For these geometries $t_p=1$ or $0$, $s=3$ and there is a $G_6$ (or in
the special case of the Einstein static universe, where $t_0=0$, a
$G_7$). This result is easy to rederive. The kinematic quantities
$\dot{u}^a,\,\omega_{ab}$ and $\sigma_{ab}$ must all be zero, and that
conversely characterizes these geometries (see (5.1a) and the
subsequent argument in \cite{Ell73}). The first derivatives of the
Riemann tensor will then involve only the expansion $\Theta$ and the
partial derivatives of $\xi$ and $p$. From the Ricci equations,
$\Theta$ is not independent of $\xi$ (this can also be obtained from
(A${}_2$), as stated in \cite{SteEll68}). Therefore $t_1=t_0$ and
either $t_0=0$ or $t_0=1$, $p=1$, and (A${}_1$) is sufficient.

The remaining spacetimes to consider are those where $s_1=1=s$. The
frame can as before be chosen so that the rotation that remains acts
in the $(\npm,\,\npmbar)$ plane.  Then $\delta \Lambda =0=\delta
\Phi_{11'}$. With that constraint, the Bianchi identities (Ba--d, g
and h), or equivalently (A${}_1$), give \eqref{DPhi01}--\eqref{DPhi23}
and \eqref{DPhiRem} (see Section \ref{CF1stDerivs}). Thus
\begin{equation}\label{CFPFBian}
\sigma=\bar{\lambda}, \quad \pi=\bar{\kappa}, \quad \tau=\bar{\nu},
  \quad \bar{\alpha}+\beta =0.
\end{equation}
In the orthonormal frame used by \cite{Ell67}, \eqref{CFPFBian}
implies $\dot{u}^2=\dot{u}^3=\omega_2=\omega_3
=\sigma_{12}=\sigma_{13}=\sigma_{23}=0$, and $\Theta_2=\Theta_3$ (so
$\sigma_{22'}=\sigma_{33'}$), as obtained by Ellis from (A${}_1$) (which
is used here just to give $\delta \Lambda =0=\delta \Phi_{11'}$).

Moreover the imaginary part of (Be) [or (Bf)] implies
\begin{equation}\label{rhomubar}
  (\rho-\bar{\rho})+(\mu-\bar{\mu})=0.
\end{equation}
which shows that $\omega_1=0$. Thus the conformally flat perfect fluids
have no vorticity (as was already known, see e.g.\ \cite{Bra86}), and
therefore belong to case II or case III of \cite{Ell67} and
\cite{SteEll68}.

% File cfpf.nul
The 4 equations (Be, f, i and k) contain combinations of the $D$ and
$\Delta$ derivatives of $\Phi_{11'}$ and $\Lambda$ of which only 3 are
linearly independent.  From (Bi)--(Be) and (Bk)--(Bf) one finds
\begin{equation}\label{consn1}
  D(\Phi_{11'}+\Lambda)= \Delta(\Phi_{11'}+\Lambda)=
[(\rho+\bar{\rho})-(\mu+\bar{\mu})]\Phi_{11'},
\end{equation}
showing that $\xi$ has a zero derivative perpendicular to the fluid
velocity.  The linear
combination with no derivative terms  (obtainable from (Be)+(Bf) using
\eqref{consn1}) is
\begin{equation}\label{lincom}
   (\mu+\bar{\mu})-(\rho+\bar{\rho})+2(\gamma+\bar{\gamma})-2(\varepsilon+\bar{\varepsilon}))=0,
\end{equation}
which shows that the ONT variable $\sigma_{11}=0$, whence also
$\sigma_{22}=0$, completing $\sigma_{ab}=0$ in agreement with
\cite{Bra86}.

Using that information and \eqref{consn1}, one obtains the
usual equation for $\partial_0\xi$ in the ONT frame.  From (Bi)--(Bk),
using \eqref{consn1}, the remaining independent combination of the
Bianchi identities (Be, f, i and k) yields
\begin{equation}\label{consn2}
  (\Delta-D)(\Phi_{11'}-3\Lambda)=
  4((\varepsilon+\bar{\varepsilon})+(\gamma+\bar{\gamma}))\Phi_{11'},
\end{equation}
  which is the usual equation relating $\partial_1p$ and $\dot{u}^1 =
  [(\varepsilon+\bar{\varepsilon})+(\gamma+\bar{\gamma})]/\sqrt{2}$ in
  the ONT frame.  Note that if $\dot{u}^1 = 0$ the conditions for
  Robertson-Walker geometry are satisfied, and $s=3$ rather than 1, so
  one must assume $\dot{u}^1 \neq 0$.

From \eqref{consn1} and \eqref{consn2} one can now compute
$(\Delta-D)\Phi_{11'}$ and $(\Delta-D)\Lambda$, which are useful in
simplifying the second derivatives.

Using \eqref{CFPFBian}, \eqref{rhomubar} and \eqref{lincom} in (NPe)
plus the conjugate of (NPd), and in (NPo) minus the conjugate of
(NPr), one obtains
\begin{equation}\label{delTheta}
  \delta(\gamma+\bar{\gamma}) = 0 =
  \delta(\varepsilon+\bar{\varepsilon}),
\end{equation}
so the ONT variables $\Theta_1$ and $\dot{u}^1$ obey
$\delta\Theta_1=0=\delta \dot{u}^1$.  Since $\sigma_{\alpha\beta}=0$,
this implies $\delta\Theta=\delta\Theta_{22}~[= \delta\Theta_{33}] =0$
also, whence $\delta(\rho+\bar{\rho})=\delta(\mu+\bar{\mu})$.
% (One can also obtain $\delta\Theta_{22}=0$ from (NPk) and the conjugate of
% (NPm).)
\cite{Ell67} used (A${}_2$) to derive the corresponding
statement, his equation (4.3e), for $\Theta_{\alpha\alpha}$ (no sum),
which is more general because in Petrov type D $\sigma_{11'}$ need not
be zero. Here \eqref{delTheta} was obtained from the consequence
$\delta \Lambda =0=\delta \Phi_{11'}$ of (A${}_1$) and the Bianchi and
Ricci identities.

Using the information obtained so far, those second derivative terms
$\nabla^2\Phi_{AB'}$ which must vanish for (A${}_2$) give
$\sigma\dot{u}^1=\kappa\dot{u}^1=\tau\dot{u}^1=0$. Since for $s=1$ one
must have $\dot{u}^1\neq 0$, one obtains \eqref{C1values}\footnote{Hence
  the ONT variables obey $n_{12}+a_3= n_{13}+a_2= n_{23}=
  \Omega_2=\Omega_3$ agreeing with results in \cite{Ell67} obtained
  from (A${}_2$).}. (NPk) gives $\delta\rho=0$ and (NPm)
$\bar{\delta}\mu=0$.  To complete the set of equations
\eqref{C1values}, \eqref{aplusb} and \eqref{deltas}, which as before imply
LRS (see Section \ref{sec:LRSD}), one needs to prove that $\delta\mu
~[=\delta\bar{\rho}]=0$. One finds
$\nabla^3\Phi_{23'}\propto
\dot{u}^1\Phi_{11'}\delta(\rho+\bar{\rho})=0$, so  (A${}_3$) gives
  $\delta\bar{\rho}=0$ as required. This agrees with (4.3f) in
  \cite{Ell67}. Hence in this case (A${}_3$) is required.

\section{Conclusion}

Summarizing the work in Sections \ref{sec:LRSD} and \ref{CFcases} gives
the following theorem

\begin{theorem} In spacetimes with a Ricci tensor of Segre type
    $[(111),1]$ whose timelike eigenvector is not geodesic, local
  spatial rotational invariance of the curvature and its derivatives
  up to the third holds if and only if the spacetime is locally
  rotationally symmetric. In all other cases, local spatial rotational
  invariance of the curvature and its first derivatives holds if and
  only if the spacetime is locally rotationally symmetric.
\end{theorem}

  \section*{Acknowledgements}
I am grateful to Jan {\AA}man for his work in developing the
software CLASSI, used above at several points where `calculation'
appears without details, and for preliminary calculations for the LRS cases,
and to Filipe Mena, working with whom on discrete isotropies prompted
this paper, and who was kind enough to check some calculations. I am
also retrospectively grateful to the late John Stewart, who first
prompted my interest in this area in 1967, leading to me joining him
as one of George Ellis's students, and to George himself.

\section*{Conflict of interests}

The author is an Editorial Board member of the journal General
Relativity and Gravitation.

\appendix
\section{Appendix: The Cahen and Defrise paper}
\label{cdapp}

This appendix explains where \cite{CahDef68} used (A${}_2$). Reference
to their equations is in the form (e.g.)\ CD(3.16). They focused on
the imaginary parts of $\mu$ and $\rho$ and took three cases.  Note
that in the following, factors of 2 or sign in the comparisons between
their formalism and the NP formalism have been neglected for brevity.

Cahen and Defrise consider 3 cases:

1. $\mu-\bar{\mu}= 0 = \rho-\bar{\rho}$ which means $\npk$ and $\npl$
are parallel to gradients. In this case they integrated the remaining
equations to obtain the metric CD(3.20).

2. $\mu-\bar{\mu} \neq 0 = \rho-\bar{\rho}$ (or the same with $\mu$
and $\rho$ swapped)\\ Applying $[\delta,\,\bar{\delta}]$ to $Q = \Lambda$,
$\Psi_2$, and $\Phi_{11'}$ one obtains $(\mu-\bar{\mu})DQ=0$ giving
CD(3.21). If $\mu=\bar{\mu}$ one is back at case 1, and if $DQ=0$ for
the three $Q$ then from (Be) and (Bi) one has
$$D\Psi_2-D\Phi_{11'}-D\Lambda/24=
3\rho\Psi_2-2\bar{\rho}\Phi_{11'}+\mu\Phi_{00'}=0.$$
Since $\rho$ is real the imaginary part of this is
$$\rho(\Psi_2-\bar{\Psi}_2)=-(\mu-\bar{\mu})\Phi_{00'}$$
(their (3.22)). Then the imaginary part of (NPq) gives
$\Psi_2-\bar{\Psi}_2=-\rho(\mu-\bar{\mu})$
so $\Phi_{00'}=2\rho^2$ giving CD(3.23).
Now (NPa) gives $D\rho=3\rho^2$ (CD(3.24)) in a frame such that
$\varepsilon+\bar{\epsilon}=0$ and then $[\delta,\,\bar{\delta}]$
applied to $\rho$ gives $\mu=\bar{\mu}$ or $D\rho=0$. Since one
assumes $\rho\neq 0$ this gives case 1 again.

3. $\mu-\bar{\mu} \neq 0 \neq \rho-\bar{\rho}.$ Here Cahen and Defrise
apply a (position-dependent) boost to make
\begin{equation}\label{murho}(\mu-\bar{\mu})=\pm
  (\rho-\bar{\rho}),
\end{equation} 
then used $[\delta,\,\bar{\delta}]$ again to get CD(3.26). The
imaginary part of (NPa) is CD(3.27) and CD(3.28) is the imaginary part
of (NPh).  They then find another expression for the imaginary part of
$\Psi_2$, which follows from (NPn) and (NPq), and compare using the
equality \eqref{murho}. CD(3.29) relates $\gamma+\bar{\gamma}$ and
$\varepsilon+\bar{\varepsilon}$. Invariance of the second derivatives
of $\Psi_2$ is used at this point to obtain $\delta\rho$ and $\bar{\delta}\mu$
(CD(3.30)), and (NPm) and (NPk) (with vanishing $\Psi_A$ other than
$\Psi_2$) to obtain the other derivatives (CD(3.31)) and then
$[\delta,\,\bar{\delta}]$ is used again to get CD(3.32)--CD(3.34). Thus one
finally arrives at CD(3.36), the equivalent of \eqref{aplusb}. As
shown above a different choice of boost enables one to show that only
(A${}_1$) is actually necessary.

% \bibliographystyle{spbasic}
%MM I prefer alphabetic sorting so one can find the reference easily,
% though I hate it when coupled with a numerical referencing style!
% \bibliographystyle{unsrtnat}
% \bibliographystyle{agsm}
% \bibliography{GRstrings,isotropy}

\end{document}